\documentclass[12pt]{article}
\usepackage{amssymb}
\newtheorem{theorem}{Theorem}[section]
\newtheorem{corollary}[theorem]{Corollary}
\newtheorem{lemma}[theorem]{Lemma}

\usepackage{amsfonts,amssymb,eucal,amsmath}
\title{\bf Saari's Conjecture for Elliptical Type $N$-Body Problem and An Application\thanks{Supported partially by NSF of China}}

\author{\small\sc Xiang Yu\footnote{Email:xiang.zhiy@gmail.com} \small{and}
\small\sc Shiqing Zhang\footnote{Email:zhangshiqing@msn.com} \\
 \small \it Department of
Mathematics, Sichuan University,
 \small\it Chengdu 610064, People's Republic of China}

\date{}

\begin{document}
\maketitle

{\bf Abstract:}  By using an arithmetic fact, we will firstly prove Saari's conjecture in a particular case, which is called the Elliptical Type N-Body Problem, and then we apply it to prove that the variational minimal solution of the planar Newtonian N-body problem is precisely a relative equilibrium solution whose
configuration minimizes the function $IU^2$, it's worth noticing that we don't need the hypothesis of Finiteness of Central Configurations. In the Planetary Restricted Problem (which ignore all the mutual gravitational interactions between the planets), the corresponding Saari's conjecture is stated and proved. \\


{\bf 2000AMS Subject Classification} 11J17, \and 11J71, \and 34C25, \and 42A16, \and 70F10, \and 70F15, \and 70G75.

\section{Introduction}
\ \ \ \  In 1970, Donald Saari \cite{saari1970bounded} proposed the following conjecture : {\it In
the Newtonian $N$-body problem, if the moment of inertia,
$I=\Sigma^n_{k=1}m_k|q_k|^2$, is constant, where $q_1,q_2,\cdots
,q_n$ represent the position vectors of the bodies of masses
$m_1,\cdots ,m_n$, then the corresponding solution is a relative
equilibrium.} In other words: Newtonian particle systems of constant
moment of inertia rotate like rigid bodies.

A lot of energies have been spent to understand Saari's conjecture, but
most of those works ( such as \cite{palmore1979relative,palmore1981saari}) failed to achieve crucial results.
However there have been a few successes in the struggle to understand
Saari's conjecture. McCord \cite{mccord2004saari} proved that the conjecture is true for
three bodies of equal masses. Llibre and Pina \cite{llibre2002saari} gave an
alternative proof of this case, but they never published it.In particular,
Moeckel \cite{moeckel2005computer,moeckel2005proof} obtained a computer-assisted proof for the Newtonian
three-body problem with positive masses when physical space
is $\mathbb{R}^d$ for all positive integer $d\geq2$. Diacu,
P$\acute{\rm e}$rez-Chavela, and Santoprete \cite{diacu2005saari} showed that the
conjectre is true for any $n$ in the collinear case for potentials
that depend only on the mutual distances between point masses. Roberts and Melanson \cite{roberts2007saari} showed that the conjecture is true
for the restricted three-body problem using a computer-assisted proof.
There have been results, such as \cite{roberts2006some,santoprete2004counterexample,schmah2007saari}, which
studied the conjecture in other contexts than the Newtonian $N$-body problem.

Recently the interest in this conjecture
has grown considerably due to the discovery of the figure eight
solution \cite{chenciner2000remarkable}, which, as numerical arguments show, has an
approximately constant moment of inertia but is not a relative
equilibrium. In recent years, for a natural extension of the original Saari's conjecture, namely Saari's homographic
conjecture, some mathematicians have made some progress \cite{diacu2008saari,fujiwara2012saaristrong,fujiwara2012saariNewton}.

The variational minimal solutions of the N-body problem are attractive, since they are nature from the viewpoint of the principle of least action.  Unfortunately, there were very few works about the variational minimal solutions before 2000.
It's worth noticing that a lot of results have been got by the action minimization methods in recent years, please see \cite{barutello2004action,chen2001action,chen2003binary,chen2008existence,chenciner2002action,Chenciner2002,chenciner1998minima,
chenciner2000remarkable,chenciner2000minima,ferrario2004existence,long2000geometric,zhang2001minimizing,zhang2002variational,zhang2004nonplanar,zhang2004new} and the references there.

Let $\mathcal{X}_d$ denote the space of configurations of $N\geq 2$ point particles with masses $m_1, \ldots,m_N$ in Euclidean space $\mathbb{R}^d$ of dimension $d$, whose center of masses is at the origin, that is, $\mathcal{X}_d = \{ q = (q_1,\cdots, q_N)\in (\mathbb{R}^d)^N: \sum_{i = 1}^{N} {m_i q_i} = 0  \}$. Let $\mathbb{T} = \mathbb{R}/T\mathbb{Z} $ denote the circle of length $T = |\mathbb{T}|$, embedded as $\mathbb{T} \subset \mathbb{R}^2$.By the loop space $\Lambda$, we mean the Sobolev space $\Lambda  = H^1(\mathbb{T}, \mathcal{X}_d)$. We consider the opposite of the potential energy (force function) defined by
\begin{equation}
U(q) = \sum_{i<j} {\frac{m_i m_j }{|q_i - q_j|}}.
\end{equation}
The kinetic energy is defined (on the tangent bundle of $\mathcal{X}_d$) by $K = \sum_{i=1}^{N} {\frac{1}{2}{m_i |\dot{q}_i|^2}}$, the total energy is $E = K- U$ and the Lagrangian is $L(q,\dot{q}) = L = K + U = \sum_i \frac{1}{2} m_i |\dot{q}|^2  + \sum_{i<j}{\frac{m_i m_j}{|q_i - q_j|}}$.
Given the Lagrangian L, the positive definite functional $\mathcal{{A}}:\Lambda \rightarrow \mathbb{R} \cup \{+\infty\}$
defined by
\begin{equation}
\mathcal{{A}}(q) = \int_{\mathbb{T}}{ L(q(t),\dot{q}(t)) dt}.
\end{equation}
 is termed as action functional (or the Lagrangian action).

The action functional $\mathcal{{A}}$ is of class $C^1$ on the subspace $\hat{\Lambda} \subset \Lambda $, which is collision-free space. Hence critical point of $\mathcal{{A}}$ in $\hat{\Lambda}$ are T-periodic classical solutions (of class $C^2$) of  Newton's equations
\begin{equation}\label{eq:Newton's equation}
m_i \ddot{q}_i = \frac{\partial U}{\partial q_i}.
\end{equation}

{\bf Definition \cite{wintner1941analytical}.} A configuration $q=(q_1,\cdots,q_N)\in {\mathcal{X}}_d\setminus\Delta_d$ is called a central configuration if there exists a constant $\lambda\in {\mathbb{R}}$ such that
\begin{equation}
\sum_{j=1,j\neq k}^N \frac{m_jm_k}{|q_j-q_k|^3}(q_j-q_k)=-\lambda m_kq_k,1\leq k\leq N
\end{equation}

The value of $\lambda$ in (1.1) is uniquely determined by
\begin{equation}
\lambda=\frac{U(q)}{I(q)}
\end{equation}

Where
\begin{equation}
\Delta_d=\left\{q=(q_1,\cdots,q_N)\in (\mathbb{R}^d)^N: q_j=q_k~\mbox{for~some}~j\neq k\right\}
\end{equation}
\begin{equation}
I(q)=\sum_{1\leq j\leq N} m_j|q_j|^2~~~~~~~~~~~~~~~~~~~
\end{equation}

It's well known that the central configurations are the critical points of the function $IU^2$, and $IU^2$ attains its infimum on ${\mathcal{X}}_d\setminus\Delta_d$. Furthermore, we know \cite{moeckel1990central} that $inf_{{\mathcal{X}}_2\setminus\Delta_2}{IU^2}<inf_{{\mathcal{X}}_1\setminus\Delta_1}{{IU^2}}$and $inf_{{\mathcal{X}}_3\setminus\Delta_3}{{IU^2}}<inf_{{\mathcal{X}}_2\setminus\Delta_2}{IU^2}$ when$N\geq4$. When $N\geq4$ and ${\mathbb{R}^d}={\mathbb{R}^3}$, it is well known that the homographic solutions derived by the central configurations minimizing the function $IU^2$  are homothetic, furthermore, a homographic motion in ${\mathbb{R}^3}$ which is not homothetic takes place in a fixed plane\cite{albouy1997probleme,arnol2006mathematical,Chenciner2002,wintner1941analytical}.This is an important reason for us only to consider $d=2$. In fact, A. Chenciner \cite{Chenciner2002} and Zhang-Zhou \cite{zhang2004nonplanar} had proved that the minimizer of Lagrangian action among (anti)symmetric loops for the spatial $N$-body($N\geq4$) problem is a collision-free non-planar solution. From the results of A. Albouy and A. Chenciner \cite{albouy1997probleme}, our idea can be applied to the case that $d$ is any positive even number, however, for the sake of simplicity, we only consider the case $d=2$. \\

The paper is structured as follows. \textbf{Section 2} introduces the Planetary Restricted Problem and
gives a precise statement of Saari's Conjecture for the Planetary Restricted Problem. \textbf{Section 3} gives our main results. \textbf{Section 4} gives the statements and proofs of some lemmas which are useful and interesting for themselves. Finally, \textbf{Section 5} gives the proofs of the main results in \textbf{Section 3} by using the lemmas in \textbf{Section 4}.

\section{Saari's Conjecture for the Planetary Restricted Problem}

The evolution of $(1+N)$-body systems (one can see \cite{chierchia2005kam}) interacting only through gravitational attraction is governed by Newton's equations (\ref{eq:Newton's equation}).
Equations (\ref{eq:Newton's equation}) are equivalent to the standard Hamilton's equations corresponding to the Hamiltonian function

 \begin{equation}
 H(p,q) =  K - U = \sum_{0\leq i\leq N} \frac{1}{2m_i} |p_i|^2  - \sum_{0\leq i<j\leq N}{\frac{m_i m_j}{|q_i - q_j|}}
\end{equation}
where $(p,q)=(p_0, \cdots, p_N;q_0, \cdots, q_N)$ are standard symplectic variables. The symplectic form is the standard one.

Introducing the symplectic coordinate change  $(p,q)=\phi_{hel}(P,Q)$:

\begin{equation}
\phi_{hel}:\begin{array}{c}
              q_0=Q_0,q_i=Q_0+Q_i  (i=1, \cdots, N) \\
             p_0=P_0- \sum_{1\leq i\leq N} P_i, p_i=P_i  (i=1, \cdots, N)
           \end{array}
\end{equation}
one sees that the new Hamiltonian $H_{hel}=H\circ\phi_{hel}$ does not depend upon $Q_0$.
This means that $P_0$ (total
linear momentum) is a global integral of motion.
Without loss of generality, one can suppose that $P_0=0$ since the invariance of the equation (\ref{eq:Newton's equation}) under the changes of inertial reference
frames.

In the ``planetary" case, one assumes that one of
the bodies, say $i = 0$ (the Sun), has mass much larger
than that of the other bodies (this accounts for the
index "hel", which stands for ``heliocentric").To
make the problem transparent, one may introduce the following rescalings. Let
$m_i=\epsilon \widetilde{m}_i, y_i=\frac{P_i}{\epsilon m^{{5}/{3}}_0}, x_i=\frac{Q_i}{m^{{2}/{3}}_0}, (i=1, \cdots, N)$,
we rescale time by a factor $\epsilon m^{{7}/{3}}_0$ (which amounts
to dividing the new Hamiltonian by such a
factor); then, the flow of the Hamiltonian function $H_{hel}$ is equivalent to the flow of the following Hamiltonian function:
 \begin{equation}
 H_{new}(y,x) = \sum_{1\leq i\leq N} (\frac{|y_i|^2}{2\mu_i}-\frac{\mu_iM_i}{|x_i|}) + \epsilon\sum_{1\leq i<j\leq N}(y_i \cdot y_j-\frac{\widetilde{m}_i\widetilde{m}_j/{m^2_0}}{|x_i - x_j|}),
\end{equation}
where the mass parameters are defined as
\begin{equation}
M_i \triangleq 1+\epsilon \frac{\widetilde{m}_i}{m_0}, ~~~~~~~~~~ \mu_i \triangleq \frac{\widetilde{m}_i}{m_0+\epsilon\widetilde{m}_i}=\frac{\widetilde{m}_i}{m_0}\frac{1}{M_i}
\end{equation}
By using these elements, the moment of inertia $I=\Sigma^N_{i=0}m_i|q_i|^2$ and force function $ U(q) = \sum_{i<j} {\frac{m_i m_j }{|q_i - q_j|}}$ can be expressed as
\begin{equation}
I=\Sigma^N_{i=0}m_i|q_i|^2= \epsilon m^{{4}/{3}}_0[\sum_{1\leq i\leq N} \widetilde{m}_i|x_i|^2-\frac{\epsilon(\sum_{1\leq i\leq N} \widetilde{m}_ix_i)^2}{\epsilon\sum_{1\leq i\leq N} \widetilde{m}_i+m_0}]
\end{equation}
\begin{equation}
U= \epsilon m^{{4}/{3}}_0[\sum_{1\leq i\leq N}\frac{\mu_iM_i }{|x_i|}+\epsilon\sum_{1\leq i<j\leq N} {\frac{\widetilde{m}_i \widetilde{m}_j/m^2_0 }{|x_i - x_j|}}]
\end{equation}
By using  rescalings, we can think that
\begin{equation}
I=\sum_{1\leq i\leq N} \widetilde{m}_i|x_i|^2-\frac{\epsilon(\sum_{1\leq i\leq N} \widetilde{m}_ix_i)^2}{\epsilon\sum_{1\leq i\leq N} \widetilde{m}_i+m_0}
\end{equation}
\begin{equation}
U= \sum_{1\leq i\leq N}\frac{\mu_iM_i }{|x_i|}+\epsilon\sum_{1\leq i<j\leq N} {\frac{\widetilde{m}_i \widetilde{m}_j/m^2_0 }{|x_i - x_j|}}
\end{equation}

For the Planetary Restricted Problem, that is the Planetary Problem when $\epsilon=0$, the Hamiltonian becomes
 \begin{equation}
 H_{0}(y,x) = \sum_{1\leq i\leq N} (\frac{|y_i|^2}{2\varrho_i}-\frac{\varrho_i}{|x_i|}),
\end{equation}
 where $\varrho_i= \frac{\widetilde{m}_i}{m_0}$.
The systems with Hamiltonian $ H_{0}$ are integrable and represent the sum of N two-body systems formed by the Sun and the $i$-th
planet (disregarding the interaction with the
other planets). In the same time, the moment of inertia $I$ and force function $U$ become
\begin{equation}
I_0=\sum_{1\leq i\leq N} \widetilde{m}_i|x_i|^2
\end{equation}
\begin{equation}
U_0= \sum_{1\leq i\leq N}\frac{\varrho_i}{|x_i|}
\end{equation}

For Two-body Problem (one can see \cite{goldstein1962classical}), Newton's equation is
\begin{equation}
\ddot{\mathbf{r}}=-\frac{\kappa\mathbf{r}}{|\mathbf{r}|^3},
\end{equation}
suppose the solution $\mathbf{r}(t)$ is ellipse,  $a$ denotes semi-major axis, $e$ denotes eccentricity, $T$ denotes  period,
$\tilde{n}=2\pi/T$ denotes mean motion, $E$ denotes eccentric anomaly, $\tau=\tilde{n}(t-\iota)$ denotes mean anomaly, where $\iota$ denotes time of perihelion passage. There are  Kepler's Third Law: $\tilde{n}^2a^3=\kappa$ and
Kepler equation: $E-e\sin E=\tau$.  Let $r=|\mathbf{r}|$,
then $r(t)=a[1-e\cos E]$, furthermore, $E (mod 2\pi)$ is periodic with period $T$.
For the Two-body Problem corresponds to the Planetary Restricted Problem
\begin{equation}
\ddot{x_i}=-\frac{x_i}{|x_i|^3},
\end{equation}
suppose the solution $x_i(t)$ is ellipse, then ${|x_i|}=a_i(1-e_i\cos E_i)$, where $E_i (mod 2\pi)$ is periodic with period $T_i$.

It is obvious that, in the Planetary Restricted Problem, if every point particle moves uniformly in circular orbit, then the moment of inertia,
$I_0=\sum_{1\leq i\leq N} \widetilde{m}_i|x_i|^2$, is constant.  In the Planetary Restricted Problem, the Saari's Conjecture
says this is  the only case: {\it if the moment of inertia,
$I_0=\sum_{1\leq i\leq N} \widetilde{m}_i|x_i|^2$, is constant, then every point particle moves uniformly in circular orbit, that is, every eccentricity
 $e_i (i=1, \cdots, N)$ must be zero}.

\section{Main Results}

The main results in this paper are the following theorems:

\begin{theorem}\label{elliptical}
Saari's Conjecture is true if $i$-th point particle has mode of motion
\begin{equation}
q_i(t)=a_i\cos(\theta(t))+b_i\sin(\theta(t)),~~~~~~\forall t\in\mathbb{T}.
\end{equation}
and $a_i, b_i\in\mathbb{R}^d$ for all $i=1,\ldots, N $, $[\varphi,\varphi+\pi]\subseteq\{\theta(t):t\in\mathbb{T}\}$ for some $\varphi\in\mathbb{R}$.
In particular, Saari's Conjecture is true when $\theta(t)=\frac{2\pi}{T}t$.
\end{theorem}

\begin{corollary}\label{centralconfigurations}
Saari's Conjecture is true if in a barycentric reference frame the configurations formed by the bodies remain
the  central configurations all the time.
\end{corollary}
{\bf Remark.}  If the Conjecture on the Finiteness of Central Configurations is true \cite{hampton2006finiteness,smale1998mathematical,wintner1941analytical}, then the {\bf Corollary \ref{centralconfigurations}} is obvious, but we don't need this hypothesis here, so the {\bf Corollary \ref{centralconfigurations}} is not trivial.\\

\begin{theorem}\label{PlanetaryRestrictedProblem}
In the Planetary Restricted Problem, the Saari's Conjecture is true.
\end{theorem}

\begin{theorem}\label{minimize}
For Newtonian N-body problem, the regular solutions minimizing the functional ${\mathcal{A}}$ in $\mathcal{S}={\{q\in H^1(\mathbb{T}, (\mathbb{R}^2)^N):\int_{\mathbb{T}}{ q(t) dt}=0}\}$ are precisely the relative equilibrium solutions whose
configurations minimize the function $IU^2$ in ${\mathbb{R}^2}$.
\end{theorem}

{\bf Remark.}  Compared with the result of A.Chenciner \cite{Chenciner2002} and Checiner-Desolneux \cite{chenciner1998minima}: For the planar $N$-body problem, a relative equilibrium solution whose
configuration minimizes $I^\frac{1}{2}U$
is always a minimizer of the action on $\mathcal{S}$; moreover, all minimizers are of this form provided there exists only a finite number
of similitude classes of $N$-body central configurations. For the second part, he could only prove rigorously for 3-body and 4-body problems, since we
know that the Conjecture on the Finiteness of Central Configurations have only been proved for 3-body and 4-body problems until now \cite{hampton2006finiteness}.

\section{Some Lemmas}

\ \ \\

Let $[t]$ denote the unique integer such that $t-1<[x]\leq t$ for any real $t$. The difference $t-[t]$ is written as $\{t\}$ and satisfies $0\leq\{t\}<1$.

First of all, we need a  famous arithmetic fact which belongs to Kronecker:

\begin{lemma}\label{Kronecker}
If 1,$\theta_1$, \ldots, $\theta_n$ are linearly independent over the rational field, then the set \{($\{k\theta_1\}$, \ldots, $\{k\theta_n\}$): $k \in \mathbb{N}\} $ are dense in the $n$-dim unite cube $\{(\varphi_1,\ldots, \varphi_n): 0\leq\varphi_i\leq1, i=1,\ldots, n\}$.
\end{lemma}

In the following, we will prove three lemmas which are needed to prove our main results, and these lemmas are also interesting for themselves.

\begin{lemma}\label{shulun}
Given $\theta_1$, \ldots, $\theta_n$ and any $\epsilon>0$, there are infinitely many integers $k\in \mathbb{N}$ such that $\{k\theta_i\}<\epsilon$ or
$\{k\theta_i\}>1-\epsilon$ for every $i=1,\ldots,n$.
\end{lemma}
{\bf Proof of Lemma \ref{shulun}:}\\

If all of $\theta_1$, \ldots, $\theta_n$ are rational, the proposition is obviously right. Hence, without loss of generality, we will suppose that 1,$\theta_1$, \ldots, $\theta_l$($1\leq l\leq n$) are linearly independent over the rational field and $\theta_{l+1}$, \ldots, $\theta_n$ can be spanned by rational linear combination, that is, we have $\theta_i=x_i^0+\sum_{1\leq j\leq l}x_i^j\theta_j$, where $l<i\leq n$ and $x_i^j$ are rational numbers for $0\leq j\leq l$. Let integer $p$ satisfy that all of $px_i^0$ are integers for $l<i$. It is easy to know that 1,$p\theta_1$, \ldots, $p\theta_l$ are still linearly independent over the rational field. Then for any $\delta>0$, there are infinitely many integers $k\in \mathbb{N}$ such that $\{kp\theta_i\}<\delta$ or
$\{kp\theta_i\}>1-\delta$ for every $i=1,\ldots,l$ by the ${\mathbf{Lemma~ \ref{Kronecker}}}$ , and it is easy to know that $\{kp\theta_i\}<C\delta$ or
$\{kp\theta_i\}>1-C\delta$  for some constant $C$ which only depends on $x_i^j$. So for any $\epsilon>0$, there are infinitely many integers $k\in \mathbb{N}$ such that $\{k\theta_i\}<\epsilon$ or
$\{k\theta_i\}>1-\epsilon$ for every $i=1,\ldots,n$.\\

$~~~~~~~~~~~~~~~~~~~~~~~~~~~~~~~~~~~~~~~~~~~~~~~~~~~~~~~~~~~~~~~~~~~~~~~~~~~~~~~~~~~~~~~~~~~~~~~~~~~~~~~~~~~~~~~~\Box$\\

\begin{lemma}\label{ellipticalconstant}
If $U(q)\equiv const$, where $q=(q_1,\cdots, q_N)$,
 \begin{equation}
q_i(t)=a_i\cos(\theta(t))+b_i\sin(\theta(t)),~~~~~~\forall t\in\mathbb{T}.
\end{equation}
and $a_i, b_i\in\mathbb{R}^d$ for all $i=1,\ldots, N $, $[\varphi,\varphi+\pi]\subseteq\{\theta(t):t\in\mathbb{T}\}$ for some $\varphi\in\mathbb{R}$. Then $q_i(t)(i=1,\ldots, N) $ is is a rigid motion.
\end{lemma}
{\bf Proof of Lemma \ref{ellipticalconstant}:}\\

Firstly, we expand $U(q(t))$ as Fourier series:
\begin{eqnarray}
U& = &\sum_{1\leq j<k\leq N} \frac{m_jm_k}{|q_j-q_k|}\nonumber\\
& = &\sum_{1\leq j<k\leq N} \frac{m_jm_k}{[|a_j-a_k|^2\cos^2\theta(t)+|b_j-b_k|^2\sin^2\theta(t)+2(a_j-a_k)\cdot(b_j-b_k)\sin\theta(t)\cos\theta(t)]^\frac{1}{2}}\nonumber\\
& = &\sum_{1\leq j<k\leq N} \frac{m_jm_k}{[\frac{|a_j-a_k|^2+|b_j-b_k|^2}{2}+(\frac{|a_j-a_k|^2-|b_j-b_k|^2}{2}) \cos(2\theta(t))+(a_j-a_k)\cdot(b_j-b_k)\sin(2\theta(t))]^\frac{1}{2}}\nonumber\\
& = &\sum_{1\leq j<k\leq N} \frac{m_jm_k}{[\frac{|a_j-a_k|^2+|b_j-b_k|^2}{2}+(\frac{|a_j-a_k|^2-|b_j-b_k|^2}{2}) \cos(2\theta(t))+(a_j-a_k)\cdot(b_j-b_k)\sin(2\theta(t))]^\frac{1}{2}}\nonumber\\
& = &\sum_{1\leq j<k\leq N} \frac{m_jm_k}{[A_{jk}+B_{jk}\cos(2\theta(t)+\theta_{jk})]^\frac{1}{2}}\nonumber
\end{eqnarray}
where
 \begin{equation}
A_{jk}=\frac{|a_j-a_k|^2+|b_j-b_k|^2}{2}
\end{equation}
\begin{equation}
B_{jk}=[(\frac{|a_j-a_k|^2-|b_j-b_k|^2}{2})^2 +((a_j-a_k)\cdot(b_j-b_k))^2]^\frac{1}{2}
\end{equation}
and $\theta_{jk}$ can be determined when $B_{jk}>0$. In the following, we will prove $B_{jk}=0$ for any $j, k \in \{{1,\ldots,N}\}$.
It is easy to know that $A_{jk}\geq B_{jk}$, let $C_{jk}=\frac{B_{jk}}{A_{jk}}$, then we have
\begin{equation}
\begin{aligned}
U & = \sum_{1\leq j<k\leq N} \frac{m_jm_k}{A_{jk}^\frac{1}{2}}[1+(-\frac{1}{2})C_{jk}\cos(2\theta(t)+\theta_{jk})
+\ldots+\\
&\frac{(-\frac{1}{2})(-\frac{1}{2}-1)\ldots(-\frac{1}{2}-n+1)}{n!}(C_{jk})^n\cos^n(2\theta(t)+\theta_{jk})+\ldots]\nonumber\\
& = \sum_{1\leq j<k\leq N} \frac{m_jm_k}{A_{jk}^\frac{1}{2}}\{1+(-\frac{1}{2})C_{jk}\frac{\exp\sqrt{-1}(2\theta(t)+\theta_{jk})+\exp-\sqrt{-1}(2\theta(t)+\theta_{jk})}{2}
 +\ldots+\\
&\frac{(-\frac{1}{2})(-\frac{1}{2}-1)\ldots(-\frac{1}{2}-n+1)}{n!}(C_{jk})^n[\frac{\exp\sqrt{-1}(2\theta(t)+\theta_{jk})+\exp-\sqrt{-1}(2\theta(t)+\theta_{jk})}{2}]^n\\
&+\ldots\}\nonumber\\
\end{aligned}
\end{equation}
\begin{equation}
\begin{aligned}
 &= \sum_{1\leq j<k\leq N} \frac{m_jm_k}{A_{jk}^\frac{1}{2}}[1+(-\frac{1}{2})C_{jk}\frac{\exp\sqrt{-1}(2\theta(t)+\theta_{jk})+\exp-\sqrt{-1}(2\theta(t)+\theta_{jk})}{2} +\ldots+\\
 &\frac{(-\frac{1}{2})(-\frac{1}{2}-1)\ldots(-\frac{1}{2}-n+1)}{n!}(C_{jk})^n\frac{\sum_{0\leq l\leq n}\left(
                                                                                                       \begin{array}{c}
                                                                                                         n \\
                                                                                                         l \\
                                                                                                       \end{array}
                                                                                                     \right)
\exp\sqrt{-1}((2\theta(t)+\theta_{jk})(2l-n))}{2^n} +\\
&\ldots]\nonumber\\
&= \sum_{1\leq j<k\leq N} \frac{m_jm_k}{A_{jk}^\frac{1}{2}}\{1+\sum_{1\leq l}\frac{(-\frac{1}{2})(-\frac{1}{2}-1)\ldots(-\frac{1}{2}-2l+1)}{(2l)!}(C_{jk})^{2l}\frac{\left(
                                                                                                       \begin{array}{c}
                                                                                                         2l \\
                                                                                                         l \\
                                                                                                       \end{array}
                                                                                                     \right)}{2^{2l}}+\\
 &\sum_{1\leq n}\exp\sqrt{-1}(2n\theta(t))[\frac{(-\frac{1}{2})(-\frac{1}{2}-1)\ldots(-\frac{1}{2}-n+1)}{n!}\frac{(C_{jk})^n\exp\sqrt{-1}(n\theta_{jk})}{2^n}+\\
&\frac{(-\frac{1}{2})(-\frac{1}{2}-1)\ldots(-\frac{1}{2}-n-1)}{(n+2)!}\frac{(C_{jk})^{n+2}\left(
                                                                                                       \begin{array}{c}
                                                                                                         n+2 \\
                                                                                                         n+1 \\
                                                                                                       \end{array}
                                                                                                     \right)\exp\sqrt{-1}(n\theta_{jk})}{2^{n+2}}+\ldots] +\\
 &\sum_{1\leq n}\exp\sqrt{-1}(-2n\theta(t))[\frac{(-\frac{1}{2})(-\frac{1}{2}-1)\ldots(-\frac{1}{2}-n+1)}{n!}\frac{(C_{jk})^n\exp\sqrt{-1}(-n\theta_{jk})}{2^n}+\\
&\frac{(-\frac{1}{2})(-\frac{1}{2}-1)\ldots(-\frac{1}{2}-n-1)}{(n+2)!}\frac{(C_{jk})^{n+2}\left(
                                                                                                       \begin{array}{c}
                                                                                                         n+2 \\
                                                                                                         n+1 \\
                                                                                                       \end{array}
                                                                                                     \right)\exp\sqrt{-1}(-n\theta_{jk})}{2^{n+2}}+\ldots]\}
\end{aligned}
\end{equation}
Since $U\equiv const$, then by the uniqueness of Fourier series we have
\begin{equation}
\begin{aligned}
&\sum_{1\leq j<k\leq N} \frac{m_jm_k}{A_{jk}^\frac{1}{2}}[\frac{(-\frac{1}{2})(-\frac{1}{2}-1)\ldots(-\frac{1}{2}-n+1)}{n!}\frac{(C_{jk})^n\exp\sqrt{-1}(n\theta_{jk})}{2^n}+\\
&\frac{(-\frac{1}{2})(-\frac{1}{2}-1)\ldots(-\frac{1}{2}-n-1)}{(n+2)!}\frac{(C_{jk})^{n+2}\left(
                                                                                                       \begin{array}{c}
                                                                                                         n+2 \\
                                                                                                         n+1 \\
                                                                                                       \end{array}
                                                                                                     \right)\exp\sqrt{-1}(n\theta_{jk})}{2^{n+2}}+\ldots]=0
\end{aligned}
\end{equation}
\begin{equation}
\begin{aligned}
&\sum_{1\leq j<k\leq N} \frac{m_jm_k}{A_{jk}^\frac{1}{2}}[\frac{(-\frac{1}{2})(-\frac{1}{2}-1)\ldots(-\frac{1}{2}-n+1)}{n!}\frac{(C_{jk})^n\exp-\sqrt{-1}(n\theta_{jk})}{2^n}+\\
&\frac{(-\frac{1}{2})(-\frac{1}{2}-1)\ldots(-\frac{1}{2}-n-1)}{(n+2)!}\frac{(C_{jk})^{n+2}\left(
                                                                                                       \begin{array}{c}
                                                                                                         n+2 \\
                                                                                                         n+1 \\
                                                                                                       \end{array}
                                                                                                     \right)\exp-\sqrt{-1}(n\theta_{jk})}{2^{n+2}}+\ldots]=0
\end{aligned}
\end{equation}
for any $n\geq1$.
Hence we have
\begin{equation}\label{eq:zhuyao}
\sum_{1\leq j<k\leq N}D^{(n)}_{jk}\exp2\pi\sqrt{-1}(n\frac{\theta_{jk}}{2\pi}) =0
\end{equation}
for any $n\geq1$, where
\begin{equation}\label{eq:convergent}
D^{(n)}_{jk}= \frac{m_jm_kC_{jk}^n}{A_{jk}^\frac{1}{2}}[1+\frac{(\frac{1}{2}+n)(\frac{1}{2}+n+1)}{(n+1)(n+2)}\frac{(C_{jk})^2\left(
                                                                                                       \begin{array}{c}
                                                                                                         n+2 \\
                                                                                                         n+1 \\
                                                                                                       \end{array}
                                                                                                     \right)}{2^2}
+\ldots]
\end{equation}
We claim that the right side of the equation (\ref{eq:convergent}) is convergent. In fact, let
\begin{equation}
\begin{aligned}
f_{jk}&=1+\frac{(\frac{1}{2}+n)(\frac{1}{2}+n+1)}{(n+1)(n+2)}\frac{(C_{jk})^2\left(
                                                                                                       \begin{array}{c}
                                                                                                         n+2 \\
                                                                                                         n+1 \\
                                                                                                       \end{array}
                                                                                                     \right)}{2^2}
+\ldots\nonumber\\
&=1+c_1(C_{jk})^2+c_2(C_{jk})^4+\ldots+c_l(C_{jk})^{2l}+\ldots
\end{aligned}
\end{equation}
where
\begin{equation}
c_l= \frac{(\frac{1}{2}+n)(\frac{1}{2}+n+1)\ldots(2l-1-\frac{1}{2}+n)(2l-\frac{1}{2}+n)}{(n+1)(n+2)\ldots(n+2l-1)(n+2l)}\frac{\left(
                                                                                                       \begin{array}{c}
                                                                                                         n+2l \\
                                                                                                         n+l \\
                                                                                                       \end{array}
                                                                                                     \right)}{2^{2l}}
\end{equation}
Then we have
\begin{equation}
\frac{c_{l+1}}{c_l}=\frac{(2l+\frac{1}{2}+n)(2l+1+\frac{1}{2}+n)}{4(l+1)(l+1+n)}
\end{equation}
\begin{equation}
\lim_{l\rightarrow\infty}\frac{c_{l+1}}{c_l}=1
\end{equation}
Hence the series of the equation (\ref{eq:convergent}) is convergent when $(C_{jk})^2<1$. Furthermore, we can prove the convergence of the series for the equation (\ref{eq:convergent}) by using Gauss' text when $(C_{jk})^2=1$. In fact, we have
\begin{equation}
\frac{c_l}{c_{l+1}}=1+\frac{\frac{n+2}{2}}{l}+\beta_l
\end{equation}
where
\begin{equation}
\beta_l=-\frac{2n^2+2n+\frac{3}{4}+\frac{(n+\frac{1}{2})(n+\frac{3}{2})(n+2)}{2l}}{4l^2+2l(n+2)+(n+\frac{1}{2})(n+\frac{3}{2})}
\end{equation}
Since $\frac{n+2}{2}>1$ and $|\beta_l|\sim\frac{c}{l^2}$, where $c$ is a constant, then it is easy to know that the series of the equation (\ref{eq:convergent}) is convergent when $C_{jk}^2=1$.\\
From ${\mathbf{Lemma~ \ref{shulun}}}$, we know there exists some $n$ such that $n\frac{\theta_{jk}}{2\pi}=k_n+\varphi_{jk}$, where $k_n$ is an integer and $-\frac{1}{4}<\varphi_{jk}<\frac{1}{4}$. Since $D^{(n)}_{jk}\geq0$, there must be $D^{(n)}_{jk}=0$ for any $j,k$ by the equation (\ref{eq:zhuyao}). So we have $C_{jk}=0$, $|q_j-q_k| \equiv \sqrt{A_{jk}}$.

Hence $q_i(t)(i=1,\ldots, N) $ is a rigid motion.\\
$~~~~~~~~~~~~~~~~~~~~~~~~~~~~~~~~~~~~~~~~~~~~~~~~~~~~~~~~~~~~~~~~~~~~~~~~~~~~~~~~~~~~~~~~~~~~~~~~~~~~~~~~~~~~~~~~\Box$

{\bf Remark.} It is easy to know that the same result is still true when the potential function is defined by
$
U(q) = \sum_{i<j} {\frac{m_i m_j }{|q_i - q_j|^\alpha}}
$ for any $\alpha>0$ and if $U(q(t))$ is a trigonometric polynomial when $i$-th point particle has the following mode of motion  \begin{equation}
q_i(t)=a_i\cos{\theta(t)}+b_i\sin{\theta(t)},~~~~~~\forall t\in\mathbb{T}.
\end{equation}
and $a_i, b_i\in\mathbb{R}^d$, for all $i=1,\ldots, N $.\\

Two numbers $t_1$ and $t_2$ are called to be linearly dependent over the rational field, if there exist two rational numbers $s_1$ and $s_2$ (at least one of them is nonvanishing) such that $t_1s_1+t_2s_2=0$. It is easy to know that  linear dependence for two numbers over the rational field is a  equivalence relation
on the set $\mathbb{R}\backslash\{0\}$. Hence we can get a partition of any  subset of $\mathbb{R}\backslash\{0\}$.
\begin{lemma}\label{periodicindependent}
Given some continuous periodic functions $u_i(t) (i\in \Lambda$, $t\in \mathbb{R})$, for the set of all the periods of $u_i(t) (i\in \Lambda)$, suppose there are only  finite  equivalence relations  according to linear dependence over the rational field, that is, there are index subsets $\Lambda_i (i=1, \cdots, n)$ such that $\bigcup^{n}_{j=1}\Lambda_j = \Lambda$ and $\Lambda_i \bigcap \Lambda_j = \emptyset (1\leq i \neq j \leq n)$, moreover, the functions $u_i(t) (i\in \Lambda_1)$ have a common period $T_1$, $\cdots$, the functions $u_i(t)$ ($i\in \Lambda_n$) have a common period $T_n$, and $T_i, T_j$ are linearly independent over the rational field for any $1\leq i, j\leq n$. If $\sum_{i\in \Lambda}u_i(t) \equiv const$, then $\sum_{i\in \Lambda_j}u_i(t) \equiv const$ for every $j\in \{1, \cdots, n\}$.
\end{lemma}
{\bf Proof of Lemma \ref{periodicindependent}:}\\

For a function $u(t)$, we define\\
$\triangle_i u \triangleq u(t-T_i)-u(t)$, $\triangle_j\triangle_i u \triangleq \triangle_iu(t-T_j)-\triangle_iu(t)$, $\triangle^k u \triangleq \triangle_k \cdots \triangle_1u$ for any $k\in \{1, \cdots, n\}$,\\
and \\
$\widetilde{\triangle}_i u \triangleq u(t+T_i)-u(t)$, $\widetilde{\triangle}_j\widetilde{\triangle}_i u \triangleq \widetilde{\triangle}_iu(t+T_j)-\widetilde{\triangle}_iu(t)$, $\widetilde{\triangle}^k u \triangleq \widetilde{\triangle}_{n-k+1} \cdots \widetilde{\triangle}_n u$ for any $k\in \{1, \cdots, n\}$.\\

From
\begin{equation}
\sum_{i\in \Lambda}u_i(t)= \sum_{1\leq j\leq n}\sum_{i\in \Lambda_j}u_i(t)\equiv const,
\end{equation}
we can get
\begin{equation}
\triangle_1 \sum_{1\leq j\leq n}\sum_{i\in \Lambda_j}u_i(t)=\triangle_1 \sum_{2\leq j\leq n}\sum_{i\in \Lambda_j}u_i(t)=0,
\end{equation}
\begin{equation}
\triangle_2\triangle_1 \sum_{2\leq j\leq n}\sum_{i\in \Lambda_j}u_i(t)=\triangle_2\triangle_1 \sum_{3\leq j\leq n}\sum_{i\in \Lambda_j}u_i(t)=
\triangle^{2}\sum_{3\leq j\leq n}\sum_{i\in \Lambda_j}u_i(t)=0,
\end{equation}
~~~~~~~~~~~~~~~~~~~~~~~~~~~~~~~~~~~~~~~~~~~~~~~~~$\cdots$\\
\begin{equation}
 \triangle^{n-1} \sum_{i\in \Lambda_n}u_i(t)=0,
\end{equation}

Then
\begin{equation}
 \int^{T_n}_0 \triangle^{n-1} \sum_{i\in \Lambda_n}u_i(t) \exp\sqrt{-1}(k\frac{2\pi}{T_n}t)dt=0,
\end{equation}
for any $k\in \mathbb{Z}\backslash \{0\}$.\\
The above equations can be changed as
\begin{eqnarray}
0 &=& \int^{T_n}_0 [\triangle^{n-2} \sum_{i\in \Lambda_n}u_i(t-T_{n-1})-\triangle^{n-2} \sum_{i\in \Lambda_n}u_i(t)] \exp\sqrt{-1}(k\frac{2\pi}{T_n}t)dt\nonumber\\
&=& \int^{T_n}_0 \triangle^{n-2} \sum_{i\in \Lambda_n}u_i(t)\widetilde{\triangle}_{n-1} \exp\sqrt{-1}(k\frac{2\pi}{T_n}t)dt\nonumber\\
&=& (\exp\sqrt{-1}(k\frac{2\pi T_{n-1}}{T_n})-1)\int^{T_n}_0 \triangle^{n-2} \sum_{i\in \Lambda_n}u_i(t) \exp\sqrt{-1}(k\frac{2\pi}{T_n}t)dt\nonumber\\
&~~~~~~&\cdots \nonumber\\
 &=& (\exp\sqrt{-1}(k\frac{2\pi T_1}{T_n})-1) \cdots (\exp\sqrt{-1}(k\frac{2\pi T_{n-1}}{T_n})-1) \nonumber\\
 &~~~~~~&    \int^{T_n}_0 \sum_{i\in \Lambda_n}u_i(t) \exp\sqrt{-1}(k\frac{2\pi}{T_n}t)dt\nonumber\\
\end{eqnarray}
for any $k\in \mathbb{Z}\backslash \{0\}$.

Since $T_n, T_j$ are linearly independent over the rational field for any $1\leq  j\leq n-1$, we can get
\begin{equation}
 \int^{T_n}_0 \sum_{i\in \Lambda_n}u_i(t) \exp\sqrt{-1}(k\frac{2\pi}{T_n}t)dt=0,
\end{equation}
for any $k\in \mathbb{Z}\backslash \{0\}$.

Hence $\sum_{i\in \Lambda_n}u_i(t) \equiv const$ holds.

Similarly, we can also get $\sum_{i\in \Lambda_j}u_i(t) \equiv const$ for every $j\in \{1, \cdots, n-1\}$.\\

$~~~~~~~~~~~~~~~~~~~~~~~~~~~~~~~~~~~~~~~~~~~~~~~~~~~~~~~~~~~~~~~~~~~~~~~~~~~~~~~~~~~~~~~~~~~~~~~~~~~~~~~~~~~~~~~~\Box$\\

\section{The Proofs of Main Results}

\ \ \\

{\bf Proof of Theorem \ref{elliptical}:}\\

From the Jacobi's identity, we known that $U$ is constant on the solution for Newtonian particle systems of constant
moment of inertia, so we can get \textbf{Theorem \ref{elliptical}} by  \textbf{Lemma \ref{ellipticalconstant}}.\\
$~~~~~~~~~~~~~~~~~~~~~~~~~~~~~~~~~~~~~~~~~~~~~~~~~~~~~~~~~~~~~~~~~~~~~~~~~~~~~~~~~~~~~~~~~~~~~~~~~~~~~~~~~~~~~~~~\Box$\\

{\bf Proof of Corollary \ref{centralconfigurations}:}\\

From the conditions of $ \mathbf{Corollary ~\ref{centralconfigurations}}$, we have
\begin{equation}
m_i \ddot{q}_i = -\lambda m_iq_i.
\end{equation}
where $\lambda=\frac{U(q)}{I(q)}$ is a constant. It is easy to know that
\begin{equation}
q_i(t)=a_i\cos(\sqrt{\lambda}t)+b_i\sin(\sqrt{\lambda}t),~~~~~~\forall t\in\mathbb{T}.
\end{equation}
for some $a_i, b_i\in\mathbb{R}^d$, $i=1,\ldots, N $.

Then by {\bf Theorem \ref{elliptical}},  we know that  the Saari's Conjecture is true.\\
$~~~~~~~~~~~~~~~~~~~~~~~~~~~~~~~~~~~~~~~~~~~~~~~~~~~~~~~~~~~~~~~~~~~~~~~~~~~~~~~~~~~~~~~~~~~~~~~~~~~~~~~~~~~~~~~~\Box$\\\\

{\bf Proof of Theorem \ref{PlanetaryRestrictedProblem}:}\\

If the solution $(x_1(t), \cdots, x_N(t))$ of the Planetary Restricted Problem satisfies $I_0=\sum_{1\leq i\leq N} \widetilde{m}_i|x_i|^2 \equiv const$,  it is easy to know that $U_0= \sum_{1\leq i\leq N}\frac{\varrho_i}{|x_i|} \equiv const$ is true. Then we know that every point particle does not collide with the sun,  otherwise, $U_0$ can not be constant since $U_0$ will tend to $\infty$ for the collision orbit; every point particle moves in elliptic orbit, otherwise, the moment of inertia $I_0$ can not be constant since $T_0$ will tend to $\infty$ for the parabolic or hyperbolic orbit.
So we have
\begin{equation}
I_0= \sum_{1\leq i\leq N} {\widetilde{m}_i}{a^2_i}(1-e_i\cos E_i)^{2}
\end{equation}
\begin{equation} \label{eq:U=const}
U_0= \sum_{1\leq i\leq N}\frac{\varrho_i}{a_i(1-e_i\cos E_i)}
\end{equation}

Our aim is to prove that every eccentricity $e_i, (i=1, \cdots, N)$ must be zero. We will mainly use the equation (\ref{eq:U=const}), it will be convenient to divide the proof into several steps.

{\bf Step 1.}

If N point particles have the same period $T$, then N point particles have the same semi-major axis $a$ by Kepler's Third Law,
their mean anomaly are respectively $\tau_i=\tilde{n}t-\tilde{n}\iota_i$. We will prove $e_i, (i=1, \cdots, N)$ must be zero in this case.

 From Kepler equation, one can get (one can see \cite{arnol2006mathematical}):
\begin{equation}
\frac{1}{1-e_i\cos E_i}=1+2\sum_{n\geq1}J_n(ne_i)\cos ({n\tau_i})
\end{equation}
where
\begin{equation}
J_n(z)=\frac{1}{2\pi}\int^{2\pi}_0 \cos (n\theta-z\sin \theta)d\theta= \sum_{k\geq0}\frac{(-1)^k(z/2)^{n+2k}}{k!(n+k)!}
\end{equation}
is the Bessel function of order $n$.

Then we have
\begin{eqnarray}
U_0 &=& \sum_{1\leq i\leq N}\frac{\varrho_i}{a(1-e_i\cos E_i)}\nonumber\\
&=& \sum_{1\leq i\leq N}\frac{\varrho_i}{a}[1+2\sum_{n\geq1}J_n(ne_i)\cos ({n\tau_i})]\nonumber\\
&=& \sum_{1\leq i\leq N}\frac{\varrho_i}{a}+\sum_{n\geq1}[\sum_{1\leq i\leq N}\frac{2\varrho_i}{a}J_n(ne_i)\cos ({n\tilde{n}\iota_i})\cos(n\tilde{n}t)\nonumber\\
&+& \sum_{1\leq i\leq N}\frac{2\varrho_i}{a}J_n(ne_i)\sin(n\tilde{n}\iota_i)\sin(n\tilde{n}t)]
\end{eqnarray}
Since $U_0 \equiv const$, we get
\begin{equation}\label{eq:yuxian}
\sum_{1\leq i\leq N}{\varrho_i}J_n(ne_i)\cos ({n\tilde{n}\iota_i})=0
\end{equation}
\begin{equation}
\sum_{1\leq i\leq N}{\varrho_i}J_n(ne_i)\sin(n\tilde{n}\iota_i)=0
\end{equation}
If $e_i>0$, then we can find the asymptotic formula for $J_n(ne_i)$ (one can see \cite{courant2008methods}):
\begin{equation}
J_n(ne_i) = \frac{2}{\sqrt{2\pi n \tanh \gamma_i}}\exp n(\tanh \gamma_i-\gamma_i)(1+{\it O}(n^{-1/5})),
\end{equation}
where $e_i=\frac{1}{\cosh \gamma_i}$ and $\gamma_i>0$,
hence $J_n(ne_i)>0$ holds for sufficiently large $n$. By \textbf{Lemma \ref{shulun}}, we know there exists some sufficiently large $n$ such that $n\tilde{n}\iota_i= 2\pi(k_{ni}+\varphi_{ni})$, where $k_{ni}$ is an integer and $-\frac{1}{4}<\varphi_{ni}<\frac{1}{4}$. Since ${\varrho_i}J_n(ne_i)>0$, we will get
\begin{equation}
\sum_{1\leq i\leq N}{\varrho_i}J_n(ne_i)\cos ({n\tilde{n}\iota_i})>0
\end{equation}
 this is a contradiction with the equation (\ref{eq:yuxian}).
 So there must be $e_i=0$ for any $i\in \{1,\cdots, N\}$ .

{\bf Step 2.}

If N point particles have different periods but they have a common period $T$. Then one can suppose that $1$-th body, $\cdots$, $N$-th body  have respectively the period  $T_1$, $\cdots$, $T_N$, and $T=k_iT_i$, where $k_i$ is positive integer,  $i\in \{1, \cdots, N\}$.

 Since
\begin{eqnarray}
U_0 &=& \sum_{1\leq i\leq N}\frac{\varrho_i}{a_i(1-e_i\cos E_i)}\nonumber\\
&=& \sum_{1\leq i\leq N}\frac{\varrho_i}{a_i}[1+2\sum_{n\geq1}J_n(ne_i)\cos ({n k_i\frac{2\pi}{T}(t-\iota_i)})]\nonumber\\
&=& \sum_{1\leq i\leq N}\frac{\varrho_i}{a_i}+\sum_{n\geq1}[\sum_{1\leq i\leq N}\frac{2\varrho_i}{a_i}J_n(ne_i)\cos ({n k_i\frac{2\pi}{T}\iota_i})\cos(n k_i\frac{2\pi}{T}t)\nonumber\\
&+& \sum_{1\leq i\leq N}\frac{2\varrho_i}{a_i}J_n(ne_i)\sin(n k_i\frac{2\pi}{T}\iota_i)\sin(n k_i\frac{2\pi}{T}t)]\nonumber\\
&=& \sum_{1\leq i\leq N}\frac{\varrho_i}{a_i}+\sum_{n\geq1}[\sum_{i\in \Sigma_n}\frac{2\varrho_i}{a_i}J_{n/k_i}(\frac{n}{k_i}e_i)\cos ({n \frac{2\pi}{T}\iota_i})\cos(n \frac{2\pi}{T}t)\nonumber\\
&+& \sum_{i\in \Sigma_n}\frac{2\varrho_i}{a_i}J_{n/k_i}(\frac{n}{k_i}e_i)\sin(n \frac{2\pi}{T}\iota_i)\sin(n \frac{2\pi}{T}t)]\nonumber\\
\end{eqnarray}
where $\Sigma_n$ is the subset of $\{1, \cdots, N\}$, whose element $i$ is a divisor of $n$.

We have
\begin{equation}\label{eq:yuxian2}
\sum_{i\in \Sigma_n}\frac{2\varrho_i}{a_i}J_{n/k_i}(\frac{n}{k_i}e_i)\cos ({n \frac{2\pi}{T}\iota_i})=0
\end{equation}
\begin{equation}
\sum_{i\in \Sigma_n}\frac{2\varrho_i}{a_i}J_{n/k_i}(\frac{n}{k_i}e_i)\sin(n \frac{2\pi}{T}\iota_i)=0
\end{equation}

Then it is similar to {\bf Step 1}, if some $e_i>0$, then we can find some sufficiently large $n$ such that
\begin{equation}
\sum_{i\in \Sigma_n}\frac{2\varrho_i}{a_i}J_{n/k_i}(\frac{n}{k_i}e_i)\cos ({n \frac{2\pi}{T}\iota_i})>0.
\end{equation}
However this result contradicts with the equation (\ref{eq:yuxian2}). So there must be $e_i=0$ for any $i\in \{1,\cdots, N\}$.

{\bf Step 3.}

If N point particles have different periods and they don't have a common period. We firstly divide these periods according to the equivalence relations  of linear dependence over the rational field. One can suppose that the family of sets $\Omega_1$, $\cdots$, $\Omega_n$ ($1\leq n\leq N$) is  the partition of these periods, and the corresponding point particles constitute respectively the sets $\Sigma_1$, $\cdots$, $\Sigma_n$ ($1\leq n\leq N$). By {\bf Lemma \ref{periodicindependent}}, we have
\begin{equation}
\sum_{i\in \Sigma_1}\frac{\varrho_i}{a_i(1-e_i\cos E_i)} \equiv const
\end{equation}
~~~~~~~~~~~~~~~~~~~~~~~~~~~~~~~~~~~~~~~~~~~~~~~~~$\cdots$\\
\begin{equation}
\sum_{i\in \Sigma_n}\frac{\varrho_i}{a_i(1-e_i\cos E_i)} \equiv const
\end{equation}

Then by {\bf Step 2},  we know that  the Saari's Conjecture is true in the Planetary Restricted Problem.\\
$~~~~~~~~~~~~~~~~~~~~~~~~~~~~~~~~~~~~~~~~~~~~~~~~~~~~~~~~~~~~~~~~~~~~~~~~~~~~~~~~~~~~~~~~~~~~~~~~~~~~~~~~~~~~~~~~\Box$\\\\

{\bf Proof of Theorem \ref{minimize}:}\\

We have
\begin{eqnarray}
{\mathcal{A}}(q)& = &\int_{\mathbb{T}}{ [\sum_i \frac{1}{2} m_i |\dot{q_i}|^2  + \sum_{i<j}{\frac{m_i m_j}{|q_i - q_j|}}] dt}\nonumber\\
& \geq &\int_{\mathbb{T}}{[(\frac{2\pi}{T})^2\sum_i \frac{1}{2} m_i |{q_i}|^2  + \sum_{i<j}{\frac{m_i m_j}{|q_i - q_j|}}] dt}\nonumber\\
& = &\int_{\mathbb{T}}{[\frac{1}{2}(\frac{2\pi}{T})^2I(q)  + \frac{1}{2}U(q)+\frac{1}{2}U(q)] dt}\nonumber\\
& \geq &3\int_{\mathbb{T}}{[(\frac{1}{2})^3(\frac{2\pi}{T})^2I(q)U^2(q)]^\frac{1}{3} dt}\nonumber\\
& \geq &3[\frac{(inf _{\mathcal{X}_2\setminus\Delta_2}{IU^2})\pi^2}{2}]^\frac{1}{3}T^\frac{1}{3}\nonumber
\end{eqnarray}
then, ${\mathcal{A}}(q)=3[\frac{(inf_{\mathcal{X}_2\setminus\Delta_2}{IU^2})\pi^2}{2}]^\frac{1}{3}T^\frac{1}{3}$ if and only if:\\
${(\textit{i})}.$ there exist $a_i, b_i\in\mathbb{R}^2$, for all $i=1,\ldots, N $, such that
 \begin{equation}
q_i(t)=a_i\cos(\frac{2\pi}{T}t)+b_i\sin(\frac{2\pi}{T}t),~~~~~~\forall t\in\mathbb{T}.
\end{equation}
${(\textit{ii})}.$
$
(\frac{2\pi}{T})^2I(q)=U(q).
$\\
${(\textit{iii})}.$  $q$ minimizes the function $IU^2$.\\
By ${(\textit{ii})}$ and ${(\textit{iii})}$ we know $I(q)\equiv const, U(q)\equiv const$, and $q(t)$ is always a central configuration.
Then
 $q$ is a relative equilibrium solution whose
configuration minimizes the function $IU^2$ by ${(\textit{i})}$ and {\bf Theorem \ref{elliptical}}.\\
$~~~~~~~~~~~~~~~~~~~~~~~~~~~~~~~~~~~~~~~~~~~~~~~~~~~~~~~~~~~~~~~~~~~~~~~~~~~~~~~~~~~~~~~~~~~~~~~~~~~~~~~~~~~~~~~~\Box$

{\bf Remark.}
We notice that as in A.Chenciner \cite{chenciner2002action} \cite{Chenciner2002} and Checiner-Desolneux \cite{chenciner1998minima}, if the Conjecture on the Finiteness of Central Configurations is true, ${(\textit{ii})}$ and ${(\textit{iii})}$ are sufficient to prove {\bf Theorem \ref{minimize}}; in fact, as \cite{chenciner2002action} pointed that if a weaker conjecture: ``the minimum points of the function $IU^2$ are finite" could be proved, ${(\textit{ii})}$ and ${(\textit{iii})}$ are also sufficient to prove {\bf Theorem \ref{minimize}}. However, we don't know any rigorous proofs for the above conjectures, hence we exploit the condition ${(\textit{i})}$ as far as possible, after we prove Saari's conjecture in the elliptical type N-Body Problem, we can get over the obstacle.\\

\section*{Acknowledgements}

 The authors sincerely thank Professor F.Diacu who told us the new progress of the Saari's conjecture.

\newpage




\end{document}